\documentclass[aps, prd, twocolumn, superscriptaddress]{revtex4-1}
\usepackage{graphicx}
\usepackage{amssymb}
\usepackage{amsmath}
\usepackage{bm}
\usepackage{subfigure}
\usepackage[colorlinks=True,linkcolor=blue,anchorcolor=blue,citecolor=blue,CJKbookmarks=true]{hyperref}

\begin{document}
\title{Vortex structures in momentum distributions from multi-pair production}

\author{Z. L. Li }
\email{Corresponding author: zlli@cumtb.edu.cn}
\affiliation{School of Science, China University of Mining and Technology, Beijing 100083, China}
\affiliation{State Key Laboratory for Tunnel Engineering, China University of Mining and Technology, Beijing 100083, China}

\author{R. Z. Jiang}
\email{Corresponding author: jiangrz2026@outlook.com}
\affiliation{State Key Laboratory for Tunnel Engineering, China University of Mining and Technology, Beijing 100083, China}

\date{\today}

\begin{abstract}
The vortex structures in the momentum distributions of multiple particle-antiparticle pairs produced from the vacuum by a spatially uniform, time-varying electric field with arbitrary polarization are investigated using the Bogoliubov transformation. For linearly polarized fields, as the field strength increases, the nodal lines in the multi-pair momentum distribution break up into vortex-antivortex pairs in both the bosonic and fermionic cases. Each vortex carries a topological charge that is an integer multiple of the single-pair vortex topological charge. For circular polarization, a central vortex appears at all field strengths. For bosons, its topological charge is an integer multiple of the single-pair vortex topological charge; for fermions, it is an integer multiple of the vortex topological charge for single-pair production with zero total spin projection along the field propagation direction. For elliptical polarization with suitable ellipticity, discrete vortices coexist with a central vortex. In this case, however, the integer-multiple relation for fermions generally fails because the two-pair production amplitude is a determinant and spin-channel contributions are separated in momentum space. These results provide a theoretical basis for understanding the vortex topology of multi-pair production in strong-field QED and for its potential control.
\end{abstract}

\maketitle

\section{INTRODUCTION}
Pair production from vacuum in strong electromagnetic fields is a long-standing prediction of quantum electrodynamics (QED) \cite{Dirac1928,Sauter1931,
Heisenberg1936,Schwinger1951,Piazza2012,Xie2017,Fedotov2023}.
For a static electric field, the threshold is set by the Schwinger critical field strength $E_{\mathrm{cr}}=m^2/e\simeq 1.3\times10^{18}\,\mathrm{V/m}$, corresponding to laser intensity $I_{\mathrm{cr}}\simeq 4.3\times10^{29}\,\mathrm{W/cm^2}$.
Here $m$ is the electron mass and $e$ is the magnitude of its charge, and natural units $\hbar=c=1$ are used throughout. Chirped-pulse amplification \cite{Strickland1985} has raised laser intensities to about  $10^{23}\,\mathrm{W/cm^2}$, still more than six orders of magnitude below $I_{\mathrm{cr}}$.
ELI facilities \cite{ELI} aim to push this limit higher, but even projected intensities remain far below $I_{\mathrm{cr}}$.
Direct experimental verification is therefore not yet feasible, and theoretical studies of pair-production signatures in complex external fields are essential to guide the design and interpretation of experiments.

For spatially homogeneous and time-dependent electric fields, the momentum distribution of produced pairs changes qualitatively with field strength.
In the subcritical regime, multi-photon absorption dominates, producing rings whose positions are fixed by energy conservation \cite{Brezin1970,Narozhnyi1974,Popov2001,Mocken2010,Abdukerim2013,
Kohlfurst2014,Sitiwaldi2017,Kohlfurst2018,Yu2023,Fan2024}. At stronger fields, tunneling comes into play, and the rings broaden and interfere, forming fringes \cite{Akkermans2012,Li2014}.
Moreover, the momentum distribution of produced pairs can exhibit spiral structures \cite{Li2017,Li2019,Hu2023,Hu2024}.
The phase of the pair-production amplitude may also exhibit singularities, corresponding to vortices that carry quantized topological charges.
Unlike the total yield, which is mainly controlled by the field intensity, the vortex structure depends sensitively on the spatiotemporal profile of the external field.

To date, studies of vortices in pair production have been largely confined to the single-pair case, where the pair-production amplitude is determined by the Bogoliubov coefficients and the momentum distribution follows from its squared modulus \cite{Bechler2023,Majczak2024,Fan2025,Jiang20262,Majczak2026,Sawut2026}.
This single-pair picture need not hold in strong fields, where multi-pair production becomes significant \cite{Popov1972,Fukushima2009,Wollert2016,Lv2017}.
Moreover, Bose-Einstein and Fermi-Dirac statistics affect multi-pair production differently.
In the simplest Bogoliubov picture, for a fixed momentum mode, bosonic multi-pair amplitudes reduce to powers of the single-pair amplitude.
For fermions, Pauli antisymmetry does not force the two-pair amplitude to vanish, since the two pairs can occupy different spin states within the same momentum mode.
The two-pair amplitude is therefore nonzero and is given by an antisymmetrized combination of the spin-resolved single-pair amplitudes.
These differences make the topology of multi-pair vortices a concrete question: do they inherit the single-pair topology, or do they form new patterns?
This cannot be answered by a mere extension of the single-pair calculation.
Note that the multi-pair production considered here means that multiple pairs are produced in the same momentum mode, not that the total number of produced pairs is greater than one.

In addition, field polarization controls the symmetry of the problem \cite{Wollert2015,Li20151,Li20152}.
Linear polarization breaks the continuous rotational symmetry about the propagation axis and gives nodal lines in the momentum distribution of produced pairs.
As the field frequency varies, these nodal lines break up into vortex-antivortex pairs \cite{Bechler2023}.
Circular polarization preserves a combined rotation-time-translation symmetry, which conserves the angular-momentum projection and can support a central vortex \cite{Fan2025}.
Elliptical polarization interpolates between these two limits and provides a continuous parameter for symmetry breaking.
So far, linear and circular polarization effects have been studied mainly in the single-pair case, whereas elliptical polarization remains largely unexplored even there.
For fermions, in particular, the spin degree of freedom may introduce additional selection rules and alter the relative weight of different channels \cite{Jiang20261}, but this has not been established in the multi-pair regime.
It remains open whether multi-pair vortex charges follow a simple scaling from the single-pair case, and whether such scaling survives broken rotational symmetry.

To address these questions, we study how field polarization and quantum statistics shape the multi-pair momentum distribution in a spatially homogeneous, time-dependent electric field.
The calculation uses the Bogoliubov transformation framework for scalar bosons and spin-1/2 fermions, with polarization varied from linear through circular to elliptical.
We compare single-pair and multi-pair results, as well as bosonic and fermionic cases under different polarizations.
The main issue is which vortex features survive from the single-pair case, which require a genuinely multi-pair description, and how these features depend on polarization and statistics.

The paper is organized as follows. Section \ref{sec:two} outlines the theoretical framework and derives the probability amplitudes for single-pair and multi-pair production. Sections \ref{sec:3a}-\ref{sec:3c} present results for linearly, circularly, and elliptically polarized fields. Section \ref{sec:four} summarizes our findings.

\section{Theoretical formula}
\label{sec:two}

\subsection{Production of multiple boson pairs}
For convenience, we use the Feshbach-Villars representation to transform the Klein-Gordon equation $(D_\mu D^\mu+m^2)\phi(\mathbf{x},t)=0$ into a Schr\"{o}dinger-type equation
\begin{equation}
i\partial_t\Phi(\mathbf{x},t)=H(\mathbf{x},t)\,\Phi(\mathbf{x},t),
\end{equation}
where $D_\mu=\partial_\mu+ieA_\mu$ is the covariant derivative, $A_\mu=(A^0,-\mathbf{A})$ is the four-vector potential of the electromagnetic field,
\begin{align}
\Phi(\mathbf{x},t)=\frac{1}{2}
\left(
  \begin{array}{c}
    \phi+\frac{i}{m}\partial_t\phi-\frac{eA^0}{m}\phi \\
    \phi-\frac{i}{m}\partial_t\phi+\frac{eA^0}{m}\phi \\
  \end{array}
\right)
\end{align}
denotes the Feshbach-Villars two-component wave function,
\begin{equation}
H(\mathbf{x},t)=\frac{(\hat{\mathbf{q}}-e\mathbf{A})^2}{2m}
(\sigma_3+i\sigma_2)+m\sigma_3+eA^0\mathbb{I}_2
\end{equation}
is the Feshbach-Villars Hamiltonian, $\hat{\mathbf{q}}=-i\bm{\nabla}$ is the canonical momentum operator, the components of $\boldsymbol{\sigma}=(\sigma_1, \sigma_2, \sigma_3)$ are the Pauli matrices, and $\mathbb{I}_2$ is a $2\times2$ identity matrix.

For a spatially homogeneous and time-dependent electric field, the canonical momentum $\mathbf{q}$ is a good quantum number.
In the temporal gauge, using the Fourier transform
\begin{equation}\label{eq:Fourier}
\Phi(\mathbf{x},t)=\int\frac{d^3q}{(2\pi)^3}e^{i\mathbf{q}
\cdot\mathbf{x}}\Phi_\mathbf{q}(t),
\end{equation}
we obtain the modular function $\Phi_\mathbf{q}(t)$ satisfying
\begin{equation}\label{eq:FVE}
i\partial_t\Phi_\mathbf{q}(t)=H_\mathbf{q}(t)\,\Phi_\mathbf{q}(t)
\end{equation}
with
\begin{equation}\label{eq:FVH}
H_{\mathbf{q}}(t)=\frac{[\mathbf{q}-e\mathbf{A}(t)]^2}{2m}
(\sigma_3+i\sigma_2)+m\sigma_3.
\end{equation}
The instantaneous eigenstates of $H_\mathbf{q}(t)$ are
\begin{eqnarray}\label{eq:InstEigen}
\begin{array}{l}
\tilde{\xi}_{\mathbf{q}}(t)=\frac{1}{\sqrt{4m\omega _{\mathbf{q}}\left( t \right) }}\left( \begin{array}{c}
	\omega _{\mathbf{q}}\left( t \right) +m\\
	-\omega _{\mathbf{q}}\left( t \right) +m\\
\end{array} \right)e^{-i\vartheta_\mathbf{q} \left( t \right)}  ,
\\
\\
\tilde{\eta}_{-\mathbf{q}}(t)=\frac{1}{\sqrt{4m\omega _{\mathbf{q}}\left( t \right) }}\left( \begin{array}{c}
	-\omega _{\mathbf{q}}\left( t \right) +m\\
	\omega _{\mathbf{q}}\left( t \right) +m\\
\end{array} \right)e^{i\vartheta_\mathbf{q} \left( t \right)} ,
\end{array}
\end{eqnarray}
where $\omega_{\mathbf{q}}(t)=\sqrt{[\mathbf{q}-e\mathbf{A}(t)]^{2}
+m^{2}}$ is the instantaneous energy and
\begin{align}\label{eq:InstPhase}
\vartheta_\mathbf{q} \left( t \right) =\int_{t_0}^t \omega_{\mathbf{q}}\left( t' \right) dt'+\omega_{\mathbf{q}}\left( t_0 \right) t_0
\end{align}
is the instantaneous phase, with $t_0$ the initial time.
The instantaneous eigenstates satisfy the eigenvalue equations
$H_{\mathbf{q}}(t)\,\tilde{\xi}_{\mathbf{q}}(t)=
\omega_{\mathbf{q}}(t)\,\tilde{\xi}_{\mathbf{q}}(t)$,
$H_{\mathbf{q}}(t)\,\tilde{\eta}_{-\mathbf{q}}(t)=
-\omega_{\mathbf{q}}(t)\,\tilde{\eta}_{-\mathbf{q}}(t)
$, the orthonormality conditions
$\tilde{\xi}_{\mathbf{q}}^{\dagger}\sigma_3\,
\tilde{\xi}_{\mathbf{q}}=1$, $\tilde{\eta}_{-\mathbf{q}}^{\dagger}\sigma_3\,
\tilde{\eta}_{-\mathbf{q}}=-1$,
$\tilde{\xi}_{\mathbf{q}}^{\dagger}\sigma_3\,
\tilde{\eta}_{-\mathbf{q}}=0$,
$\tilde{\eta}_{-\mathbf{q}}^{\dagger}\sigma_3\,
\tilde{\xi}_{\mathbf{q}}=0$,
and the completeness relation
$\tilde{\xi}_{\mathbf{q}}\tilde{\xi}_{\mathbf{q}}^{\dagger}
\sigma_3-\tilde{\eta}_{-\mathbf{q}}
\tilde{\eta}_{-\mathbf{q}}^{\dagger}\sigma_3 = \mathbb{I}_2.$

The quantized field $\hat{\Phi}$ can be expanded in the instantaneous basis as
\begin{align}\label{eq:InstBasis}
\hat{\Phi}(\mathbf{x},t)=\!\!\int\!\!\frac{d^{3}q}{(2\pi)^{3}}\,
e^{i\mathbf{q}\cdot\mathbf{x}}\bigl[\hat{a}_{\mathbf{q}}(t)\,
\tilde{\xi}_{\mathbf{q}}(t)+\hat{b}_{-\mathbf{q}}^{\dagger}(t)\,
\tilde{\eta}_{-\mathbf{q}}(t)\bigr],
\end{align}
where the time-dependent annihilation and creation operators satisfy the standard equal-time canonical commutation relations.
The instantaneous vacuum $|0_t\rangle$ is defined as the state such that $\hat{a}_{\mathbf{q}}(t)|0_t\rangle=\hat{b}_{-\mathbf{q}}(t)
|0_t\rangle=0$.
$\hat{\Phi}$ can also be expanded in the exact positive- and negative-frequency solutions $\xi_{\mathbf{q}}(t)$,  $\eta_{-\mathbf{q}}(t)$ of the time-dependent Feshbach-Villars equation (\ref{eq:FVE}):
\begin{align}\label{eq:ExactBasis}
\hat{\Phi}(\mathbf{x},t)=\!\!\int\!\!\frac{d^{3}q}{(2\pi)^{3}}\,
e^{i\mathbf{q}\cdot\mathbf{x}}\bigl[\hat{a}_{\mathbf{q}}\,
\xi_{\mathbf{q}}(t)+\hat{b}_{-\mathbf{q}}^{\dagger}\,
\eta_{-\mathbf{q}}(t)\bigr],
\end{align}
where the time-independent annihilation and creation operators satisfy the standard equal-time canonical commutation relations.
The initial vacuum $|0_{t_0}\rangle$ is defined as the state such that $\hat{a}_{\mathbf{q}}|0_{t_0}\rangle=\hat{b}_{-\mathbf{q}}(t)
|0_{t_0}\rangle=0$.
The exact solutions are selected by initial conditions at a reference time $t_{0}$:
\begin{gather}\label{eq:InitialCond}
\xi_{\mathbf{q}}(t_{0})=\tilde{\xi}_{\mathbf{q}}(t_{0}), \qquad
\eta_{-\mathbf{q}}(t_{0})=\tilde{\eta}_{-\mathbf{q}}(t_{0}).
\end{gather}
At an arbitrary time $t$ the exact solutions can be expanded in the instantaneous basis:
\begin{equation}\label{eq:Relation1}
\begin{gathered}
\xi_{\mathbf{q}}(t)=\alpha_{\mathbf{q}}(t)\,\tilde{\xi}_{\mathbf{q}}(t)
+\beta_{\mathbf{q}}(t)\,\tilde{\eta}_{-\mathbf{q}}(t),\\
\eta_{-\mathbf{q}}(t)=\gamma_{\mathbf{q}}(t)\,\tilde{\xi}_{\mathbf{q}}(t)
+\delta_{\mathbf{q}}(t)\,\tilde{\eta}_{-\mathbf{q}}(t),
\end{gathered}
\end{equation}
where $\alpha_{\mathbf{q}}(t),\beta_{\mathbf{q}}(t),
\gamma_{\mathbf{q}}(t),\delta_{\mathbf{q}}(t)$ are the Bogoliubov coefficients.
Using the Feshbach-Villars inner product, the coefficients are extracted as
\begin{equation}\label{eq:Coeff1}
\begin{gathered}
\alpha_{\mathbf{q}}(t)= \tilde{\xi}_{\mathbf{q}}^{\dagger}(t)\sigma_3\xi_{\mathbf{q}}(t), \;
\beta_{\mathbf{q}}(t)= -\tilde{\eta}_{-\mathbf{q}}^{\dagger}(t)\sigma_3\xi_{\mathbf{q}}(t),\\
\gamma_{\mathbf{q}}(t)= \tilde{\xi}_{\mathbf{q}}^{\dagger}(t)\sigma_3\eta_{-\mathbf{q}}(t), \;
\delta_{\mathbf{q}}(t)= -\tilde{\eta}_{-\mathbf{q}}^{\dagger}(t)\sigma_3\eta_{-\mathbf{q}}(t) .
\end{gathered}
\end{equation}
The initial conditions (\ref{eq:InitialCond}) give $\alpha_\mathbf{q}(t_{0})=\delta_\mathbf{q}(t_{0})=1$, $\beta_\mathbf{q}(t_{0})=\gamma_\mathbf{q}(t_{0})=0$.
The invariance of the inner product for exact solutions implies the relations
\begin{equation}
\begin{gathered}
|\alpha_{\mathbf{q}}(t)|^{2}-|\beta_{\mathbf{q}}(t)|^{2}=1,\; |\delta_{\mathbf{q}}(t)|^{2}-|\gamma_{\mathbf{q}}(t)|^{2}=1,\\ \alpha^{*}_{\mathbf{q}}(t)\gamma_{\mathbf{q}}(t)
-\beta^{*}_{\mathbf{q}}(t)\delta_{\mathbf{q}}(t)=0.%\\
%|\alpha|^{2}-|\gamma|^{2}&=1,\; |\delta|^{2}-|\beta|^{2}=1,\; \alpha\beta^{*}-\gamma\delta^{*}=0 .
\end{gathered}
\end{equation}
From the charge conjugation symmetry of bosons in the Feshbach-Villars representation, it follows that
\begin{align}
\gamma_{\mathbf{q}}(t)=\beta^{*}_{\mathbf{q}}(t),\quad \delta_{\mathbf{q}}(t)=\alpha^{*}_{\mathbf{q}}(t).
\end{align}
Substituting Eqs. (\ref{eq:Relation1}) into Eq. (\ref{eq:ExactBasis}) and comparing with the instantaneous expansion (\ref{eq:InstBasis}) yields the operator Bogoliubov transformation
\begin{equation}\label{eq:BogMat}
\begin{pmatrix}\hat{a}_{\mathbf{q}}(t)\\[2pt]
\hat{b}_{-\mathbf{q}}^{\dagger}(t)\end{pmatrix}
=
\begin{pmatrix}\alpha_{\mathbf{q}}(t)&\beta^*_{\mathbf{q}}(t)\\[2pt]
\beta_{\mathbf{q}}(t)&\alpha^*_{\mathbf{q}}(t)\end{pmatrix}
\begin{pmatrix}\hat{a}_{\mathbf{q}}\\[2pt]
\hat{b}_{-\mathbf{q}}^{\dagger}\end{pmatrix}.
\end{equation}
The inverse transformation follows from symplectic inversion,
\begin{equation}\label{eq:BogMatI}
\begin{pmatrix}\hat{a}_{\mathbf{q}}\\[2pt]
\hat{b}_{-\mathbf{q}}^{\dagger}\end{pmatrix}
=
\begin{pmatrix}\alpha_{\mathbf{q}}^{*}(t)&-\beta_{\mathbf{q}}^{*}(t)\\[2pt]
-\beta_{\mathbf{q}}(t)&\alpha_{\mathbf{q}}(t)\end{pmatrix}
\begin{pmatrix}\hat{a}_{\mathbf{q}}(t)\\[2pt]
\hat{b}_{-\mathbf{q}}^{\dagger}(t)\end{pmatrix},
\end{equation}

Applying the inverse transformation (\ref{eq:BogMatI}) to the condition $\hat{a}_{\mathbf{q}}|0_{t_{0}}\rangle=0$ gives
\begin{equation}
\hat{a}_{\mathbf{q}}(t)|0_{t_{0}}\rangle=\frac{\beta^*_{\mathbf{q}}(t)}
{\alpha^*_{\mathbf{q}}(t)}\,\hat{b}_{-\mathbf{q}}^{\dagger}(t)\,|0_{t_{0}}\rangle,
\end{equation}
which is solved by a bosonic squeezed state.
The relation between the initial vacuum and the instantaneous vacuum for a single momentum mode $\mathbf{q}$ is
\begin{equation}\label{eq:InitVacu}
|0_{t_{0}}\rangle_{\mathbf{q}}
=N_{\mathbf{q}}(t)\,
\exp \Bigl(C_{\mathbf{q}}(t)\,a_{\mathbf{q}}^{\dagger}(t)
b_{-\mathbf{q}}^{\dagger}(t)\Bigr)|0_{t}\rangle_{\mathbf{q}},
\end{equation}
where
\begin{equation}
N_{\mathbf{q}}(t)=\frac{1}{\alpha_{\mathbf{q}}(t)}, \quad C_{\mathbf{q}}(t)=\frac{\beta^*_{\mathbf{q}}(t)}{\alpha^*_{\mathbf{q}}(t)}.
\end{equation}
The normalization follows from $\langle0_{t_{0}}|0_{t_{0}}\rangle=1$ together with $|\alpha_{\mathbf{q}}(t)|^{2}-|\beta_{\mathbf{q}}(t)|^{2}=1$.
For convenience, the phase factor between the initial vacuum $|0_{t_0}\rangle_{\mathbf{q}}$ and the instantaneous vacuum $|0_t\rangle_{\mathbf{q}}$ is chosen to be $e^{-i\arg(\alpha_{\mathbf{q}}(t))}$.
The expectation value of the instantaneous particle number operator $a_{\mathbf{q}}^{\dagger}(t)a_{\mathbf{q}}(t)$ in the initial vacuum is then
\begin{equation}\label{eq:MS1}
f_\mathbf{q}(t)=\langle0_{t_{0}}|a_{\mathbf{q}}^{\dagger}(t)
a_{\mathbf{q}}(t)|0_{t_{0}}\rangle= |\beta_{\mathbf{q}}(t)|^{2},
\end{equation}
which denotes the momentum distribution function of produced bosons.
%Similarly, the antiparticle number expectation value is $|\beta_{\mathbf{q}}|^{2}$.
Expanding the exponential in Eq. (\ref{eq:InitVacu}) and using canonical commutation relations yields the multi-pair structure characteristic of bosons:
\begin{equation}
|0_{t_{0}}\rangle_{\mathbf{q}}=\frac{1}{\alpha_\mathbf{q}(t)}
\sum_{n=0}^{\infty}\frac{1}{n!}\!\left( \frac{\beta^*_\mathbf{q}(t)}
{\alpha^*_\mathbf{q}(t)}\,a_{\mathbf{q}}^{\dagger}(t)
b_{-\mathbf{q}}^{\dagger}(t)\right)^{\!\!n}|0_{t}\rangle_{\mathbf{q}}.
\end{equation}
Therefore, the probability amplitude for the creation of exactly $n$ particle-antiparticle pairs with momentum $\pm\mathbf{q}$ is
\begin{align}\label{eq:ProbAmpl1}
c_\mathbf{q}^{(n)}(t)=\frac{1}{\alpha_\mathbf{q}(t)}
\left( \frac{\beta^*_\mathbf{q}(t)}
{\alpha^*_\mathbf{q}(t)}\right)^{\!\!n},\quad n=0,1,2,\dots
%\\=&\frac{1}{|\delta_\mathbf{q}(t)|}\left(\frac{\gamma_\mathbf{q}(t)}
%{\delta_\mathbf{q}(t)}\right)^{\!\!n},\quad n=0,1,2,\dots
\end{align}
The probability is
\begin{equation}\label{eq:Prob1}
P_\mathbf{q}^{(n)}(t) =|c^{(n)}_{\mathbf{q}}(t)|^2= \frac{|\beta_{\mathbf{q}}(t)|^{2n}}{|\alpha_{\mathbf{q}}(t)|^{2n+2}}
,\;\; n=0,1,2,\dots
\end{equation}
Total probability is automatically conserved:
\begin{equation}
\sum_{n=0}^{\infty} P^{(n)}_{\mathbf{q}}(t)
= \frac{1}{|\alpha_{\mathbf{q}}(t)|^{2}}
\frac{1}{1-|\beta_{\mathbf{q}}(t)|^{2}/|\alpha_{\mathbf{q}}(t)|^{2}} = 1.
\end{equation}
The mean number of produced bosons in a single momentum mode $\mathbf{q}$ is
\begin{equation}
\sum_{n=0}^{\infty} nP^{(n)}_{\mathbf{q}}(t)
=|\beta_\mathbf{q}(t)|^{2},
\end{equation}
which is precisely the momentum distribution function.
The relative probability amplitude for produced pairs is defined as the probability amplitude for pair production divided by the vacuum-to-vacuum amplitude $\langle0_{t}|0_{t_0}\rangle_\mathbf{q}=1/\alpha_\mathbf{q}(t)$:
\begin{align}\label{eq:RelaProbA}
\widetilde{c}_\mathbf{q}^{\,(n)}(t)=
\left( \frac{\beta^*_\mathbf{q}(t)}
{\alpha^*_\mathbf{q}(t)}\right)^{\!\!n},\quad n=0,1,2,\dots
%\\=&\frac{1}{|\delta_\mathbf{q}(t)|}\left(\frac{\gamma_\mathbf{q}(t)}
%{\delta_\mathbf{q}(t)}\right)^{\!\!n},\quad n=0,1,2,\dots
\end{align}
The relative probability is
\begin{equation}\label{eq:RelaProb}
\widetilde{P}_\mathbf{q}^{\,(n)}(t) =|\widetilde{c}^{\,(n)}_{\mathbf{q}}(t)|^2= \left|\frac{\beta_{\mathbf{q}}(t)}{\alpha_{\mathbf{q}}(t)}\right|^{2n}
,\;\; n=0,1,2,\dots
\end{equation}
%Vacuum persistence amplitude
%The overlap between the instantaneous and initial vacua is the product over all modes:
%\begin{equation}
%\langle0_{t}|0_{t_{0}}\rangle =\prod_{\mathbf{q}}\mathcal{N}_{\mathbf{q}}(t)= \prod_{\mathbf{q}}\frac{1}{|\delta_{\mathbf{q}}(t)|}.
%\begin{end}
%Its squared modulus $\prod_{\mathbf{q}}|\delta_{\mathbf{q}}|^{-2}$ gives the probability that no pairs are produced throughout the evolution.

From Eqs. (\ref{eq:FVE})-(\ref{eq:InstPhase}) and Eq. (\ref{eq:InitialCond}), one can obtain the exact solutions $\xi_\mathbf{q}(t)$ and $\eta_{-\mathbf{q}}(t)$, and then, via Eqs. (\ref{eq:InstEigen}) and (\ref{eq:Coeff1}), the Bogoliubov coefficients $\alpha_\mathbf{q}(t)$ and $\beta_\mathbf{q}(t)$.
Finally, using Eqs. (\ref{eq:MS1}), (\ref{eq:ProbAmpl1}), (\ref{eq:Prob1}), (\ref{eq:RelaProbA}) and (\ref{eq:RelaProb}), we can compute the corresponding momentum distribution, probability amplitudes and probabilities.

\subsection{Production of multiple fermion pairs}

We extend the formalism developed for bosons to the case of fermions.
After Fourier decomposition of the Dirac field $\Psi(\mathbf{x}, t)$, the mode functions $\Psi_\mathbf{q}(t)$ satisfy the equation
\begin{equation}\label{eq:DE}
i\partial_t \Psi_{\mathbf{q}}(t) = \mathcal{H}_{\mathbf{q}}(t) \Psi_{\mathbf{q}}(t),
\end{equation}
where
\begin{equation}\label{eq:DEHami}
\mathcal{H}_{\mathbf{q}}(t)=\boldsymbol{\alpha}\cdot\left[\mathbf{q}
-e\mathbf{A}(t)\right]+\beta m
\end{equation}
is the Hamiltonian in momentum space, $\boldsymbol{\alpha}=(\alpha_1, \alpha_2, \alpha_3)$ and $\beta$ are Dirac alpha and beta matrices.
The instantaneous eigenstates of $\mathcal{H}_\mathbf{q}(t)$ are
\begin{equation}\label{eq:DEInst}
\begin{split}
&\tilde{u}_{\mathbf{q},s}(t) = \sqrt{\frac{\omega_{\mathbf{q}}(t) + m}{2\omega_{\mathbf{q}}(t)}}
\begin{pmatrix}
    \chi_s \\[1.2ex]
    \dfrac{\boldsymbol{\sigma} \cdot \mathbf{p}}{\omega_{\mathbf{q}}(t) + m} \chi_s
\end{pmatrix}e^{-i\vartheta_\mathbf{q} \left( t \right)}, \\[2.5ex]
&\tilde{v}_{-\mathbf{q},s}(t) = \sqrt{\frac{\omega_{\mathbf{q}}(t) + m}{2\omega_{\mathbf{q}}(t)}}
\begin{pmatrix}
    -\dfrac{\boldsymbol{\sigma} \cdot \mathbf{p}}{\omega_{\mathbf{q}}(t) + m}  \chi_s \\[1.2ex]
     \chi_s
\end{pmatrix}e^{i\vartheta_\mathbf{q} \left( t \right)},
\end{split}
\end{equation}
where $\omega_\mathbf{q}(t)$ and $\vartheta_\mathbf{q}(t)$ are the same as in the bosonic case, and $\mathbf{p}=\mathbf{q}-e\mathbf{A}(t)$ is the kinetic momentum.
The index $s=1,2$ labels spin up and spin down along the $z$ direction, respectively, with the corresponding spinors $\chi_1=(1,0)^{\intercal}$ and $\chi_2=(0,1)^{\intercal}$.
The instantaneous eigenstates satisfy the eigenvalue equations
$\mathcal{H}_{\mathbf{q}}(t)\,\tilde{u}_{\mathbf{q},s}(t)
=\omega_{\mathbf{q}}(t)\,\tilde{u}_{\mathbf{q},s}(t),\; \mathcal{H}_{\mathbf{q}}(t)\,\tilde{v}_{-\mathbf{q},s}(t)=
-\omega_{\mathbf{q}}(t)\,\tilde{v}_{-\mathbf{q},s}(t)$,
the orthonormality conditions
\begin{equation}
\begin{gathered}
\tilde{u}_{\mathbf{q},r}^{\dagger}(t)\tilde{u}_{\mathbf{q},s}(t)
=\delta_{rs},\quad
\tilde{v}_{-\mathbf{q},r}^{\dagger}(t)\tilde{v}_{-\mathbf{q},s}(t)
=\delta_{rs},\\ \tilde{u}_{\mathbf{q},r}^{\dagger}(t)\tilde{v}_{-\mathbf{q},s}(t)
=0, \quad
\tilde{v}_{-\mathbf{q},r}^{\dagger}(t)\tilde{u}_{\mathbf{q},s}(t)=0,
\end{gathered}
\end{equation}
and the completeness relation
\begin{equation}
\sum_{s=1}^2\big[\tilde{u}_{\mathbf{q},s}(t)
\tilde{u}_{\mathbf{q},s}^{\dagger}(t)
+\tilde{v}_{-\mathbf{q},s}(t)\tilde{v}_{-\mathbf{q},s}^{\dagger}(t)
\big]=\mathbb{I}_4.
\end{equation}

Similar to the bosonic case, the quantized field $\hat{\Psi}(\mathbf{x},t)$ can be expanded in both the instantaneous basis and the exact solution basis as
\begin{align}
\hat{\Psi}(\mathbf{x},t)=\int\frac{d^{3}q}{(2\pi)^{3}}e^{i\mathbf{q}
\cdot\mathbf{x}}\sum_{s=1}^2&\big[\hat{b}_{\mathbf{q},s}(t)\,
\tilde{u}_{\mathbf{q},s}(t) \nonumber\\
&+\hat{d}_{-\mathbf{q},s}^{\dagger}(t)\,
\tilde{v}_{-\mathbf{q},s}(t)\big] \label{eq:InstExpand}\\
=\int\frac{d^{3}q}{(2\pi)^{3}}e^{i\mathbf{q}\cdot\mathbf{x}}
\sum_{s=1}^2&\big[\hat{b}_{\mathbf{q},s}\,u_{\mathbf{q},s}(t) \nonumber\\
&+\hat{d}_{-\mathbf{q},s}^{\dagger}\,v_{-\mathbf{q},s}(t)\big], \label{eq:ExacExpand}
\end{align}
where both the time-dependent and time-independent annihilation and creation operators satisfy the standard equal-time canonical anticommutation relations.
The instantaneous vacuum $|0_t\rangle$ is defined by $\hat{b}_{\mathbf{q},s}(t)|0_t\rangle=\hat{d}_{-\mathbf{q},s}(t)
|0_t\rangle=0$.
The initial vacuum $|0_{t_0}\rangle$ is defined by $\hat{b}_{\mathbf{q},s}|0_{t_0}\rangle=\hat{d}_{-\mathbf{q},s}
|0_{t_0}\rangle=0$.
The exact solutions $u_{\mathbf{q},s}(t)$ and $v_{-\mathbf{q},s}(t)$ are selected by initial conditions at a reference time $t_{0}$:
\begin{gather}\label{eq:InitialCond2}
u_{\mathbf{q},s}(t_{0})=\tilde{u}_{\mathbf{q},s}(t_{0}), \qquad
v_{-\mathbf{q},s}(t_{0})=\tilde{v}_{-\mathbf{q},s}(t_{0}).
\end{gather}
At an arbitrary time $t$ the exact solutions can be expanded in the instantaneous basis:
\begin{align}
u_{\mathbf{q},s}(t)=\sum_{s'=1}^{2}
&\big[\alpha_{\mathbf{q},ss'}(t)\,\tilde{u}_{\mathbf{q},s'}(t) \nonumber \\
&+\beta_{\mathbf{q},ss'}(t)\,\tilde{v}_{-\mathbf{q},s'}(t)\big],
\label{eq:Relation2a}\\
v_{-\mathbf{q},s}(t)=\sum_{s'=1}^{2}
&\big[\gamma_{\mathbf{q},ss'}(t)\,\tilde{u}_{\mathbf{q},s'}(t) \nonumber \\
&+\delta_{\mathbf{q},ss'}(t)\,\tilde{v}_{-\mathbf{q},s'}(t)\big]. \label{eq:Relation2b}
\end{align}
where $\alpha,\beta,\gamma,\delta$ are four $2\times2$ Bogoliubov coefficient matrices for the mode functions.
By virtue of the orthonormality of the instantaneous spinors, the Bogoliubov coefficients are extracted as
\begin{equation}\label{eq:Coeff2}
\begin{gathered}
\alpha_{\mathbf{q},ss'}(t)=\tilde{u}_{\mathbf{q},s'}^{\dagger}(t)
u_{\mathbf{q},s}(t), \\
\beta_{\mathbf{q},ss'}(t)=
\tilde{v}_{-\mathbf{q},s'}^{\dagger}(t)u_{\mathbf{q},s}(t),\\
\gamma_{\mathbf{q},ss'}(t)=\tilde{u}_{\mathbf{q},s'}^{\dagger}(t)
v_{-\mathbf{q},s}(t), \\
\delta_{\mathbf{q},ss'}(t)=
\tilde{v}_{-\mathbf{q},s'}^{\dagger}(t)v_{-\mathbf{q},s}(t).
\end{gathered}
\end{equation}
The initial conditions are
\begin{equation}
\alpha(t_{0})=\mathbb{I}_2,\;\;\beta(t_{0})=0,\;\;\gamma(t_{0})=0,\;\;
\delta(t_{0})=\mathbb{I}_2.
\end{equation}
The time invariance of the Dirac inner product imposes the following unitary constraints on these matrices:
\begin{align}
\alpha\alpha^{\dagger}+\beta\beta^{\dagger}&=\mathbb{I}_2,\;\;
\gamma\gamma^{\dagger}+\delta\delta^{\dagger}=\mathbb{I}_2,\;\;
\alpha\gamma^{\dagger}+\beta\delta^{\dagger}=0,\\
\alpha^{\dagger}\alpha+\gamma^{\dagger}\gamma&=\mathbb{I}_2,\;\;
\beta^{\dagger}\beta+\delta^{\dagger}\delta=\mathbb{I}_2,\;\;
\alpha^{\dagger}\beta+\gamma^{\dagger}\delta=0.
\end{align}
%A direct consequence is $|\det\alpha|=|\det\delta|$, which will be employed in the subsequent derivations.
Substituting Eqs. (\ref{eq:Relation2a}) and (\ref{eq:Relation2b}) into Eq. (\ref{eq:ExacExpand}) and comparing with Eq. (\ref{eq:InstExpand}) yields the operator-level Bogoliubov transformations
\begin{gather}
\hat{b}_{\mathbf{q},s}(t)=\sum_{s'=1}^{2}\bigl[\alpha_{\mathbf{q},s's}
(t)\,\hat{b}_{\mathbf{q},s'}+\gamma_{\mathbf{q},s's}(t)\,
\hat{d}_{-\mathbf{q},s'}^{\dagger}\bigr],\\
\hat{d}_{-\mathbf{q},s}^{\dagger}(t)=\sum_{s'=1}^{2}
\bigl[\beta_{\mathbf{q},s's}(t)\,\hat{b}_{\mathbf{q},s'}
+\delta_{\mathbf{q},s's}(t)\,\hat{d}_{-\mathbf{q},s'}^{\dagger}\bigr].
\end{gather}
The inverse transformation is
\begin{gather}
\hat{b}_{\mathbf{q}, s} = \sum_{s'=1}^2 \big[ \alpha^*_{\mathbf{q}, s s'}(t) \, \hat{b}_{\mathbf{q}, s'}(t) + \beta^*_{\mathbf{q}, s s'}(t) \, \hat{d}^\dagger_{-\mathbf{q}, s'}(t) \big], \\
\hat{d}^\dagger_{-\mathbf{q}, s} = \sum_{s'=1}^2 \big[ \gamma^*_{\mathbf{q}, s s'}(t) \, \hat{b}_{\mathbf{q}, s'}(t) + \delta^*_{\mathbf{q}, s s'}(t) \, \hat{d}^\dagger_{-\mathbf{q}, s'}(t) \big].
\end{gather}

Imposing $\hat{b}_{\mathbf{q},s}|0_{t_{0}}\rangle=0$ gives
\begin{gather}
\sum_{s'=1}^2 \!\big[ \alpha^*_{\mathbf{q}, s s'}(t) \, \hat{b}_{\mathbf{q}, s'}(t) \!+\! \beta^*_{\mathbf{q}, s s'}(t) \, \hat{d}^\dagger_{-\mathbf{q}, s'}(t) \big]
|0_{t_{0}}\rangle=0.
\end{gather}
For a fixed momentum mode $\mathbf{q}$, we write the solution of this linear operator equation as
\begin{align}\label{eq:VacuSolu}
|0_{t_{0}}\rangle_{\mathbf{q}}\!=\mathcal{N}_{\mathbf{q}}(t)
\exp\!\left(\sum_{s,s'=1}^{2}\!\mathcal{C}_{\mathbf{q},ss'}(t)\,
\hat{b}_{\mathbf{q},s}^{\dagger}(t)\hat{d}_{-\mathbf{q},s'}^{\dagger}
(t)\!\right)\!|0_{t}\rangle_{\mathbf{q}}.
\end{align}
where
\begin{align}
\mathcal{N}_\mathbf{q}(t)=\det\alpha_\mathbf{q}(t), \quad \mathcal{C}_\mathbf{q}(t)=-\bigl[\alpha_\mathbf{q}(t)^{-1}
\beta_\mathbf{q}(t)]^*.
\end{align}
The normalization factor follows from $\langle0_{t_{0}}|0_{t_{0}}\rangle=1$ and the matrix $\mathcal{C}_\mathbf{q}(t)$ is fixed by the condition $\alpha_\mathbf{q}^*(t)\mathcal{C}_\mathbf{q}(t)
+\beta_\mathbf{q}^*(t)=0$.
For convenience, the phase factor between the initial vacuum $|0_{t_0}\rangle_{\mathbf{q}}$ and the instantaneous vacuum $|0_t\rangle_{\mathbf{q}}$ is chosen to be $e^{i\arg(\det \alpha_\mathbf{q}(t))}$.
The expectation value of the instantaneous particle number operator $\hat{b}_{\mathbf{q},s}^{\dagger}(t)\hat{b}_{\mathbf{q},s}(t)$ in the initial vacuum is then
\begin{align}\label{eq:MSF}
f_{\mathbf{q},s}(t)=&\langle0_{t_{0}}|
\hat{b}_{\mathbf{q},s}^{\dagger}(t)\hat{b}_{\mathbf{q},s}(t)
|0_{t_{0}}\rangle=\sum_{s'=1}^{2}\bigl|\gamma_{\mathbf{q},s's}(t)
\bigr|^{2} \nonumber\\
=&1-[\alpha_\mathbf{q}(t)^\dagger\alpha_\mathbf{q}(t)]_{ss},
\end{align}
which is the momentum distribution function of produced fermions.
%Similarly, the expectation value of the antiparticle number is $\sum_{s}\bigl|\beta_{\mathbf{q},s's}\bigr|^{2}$.
Expanding the exponential operator in Eq. (\ref{eq:VacuSolu}) and taking into account the Pauli exclusion principle, one finds that at most two fermion pairs can be produced for a given $\mathbf{q}$:
\begin{align}
|0_{t_{0}}\rangle_{\mathbf{q}}=&\det\alpha_\mathbf{q}(t)
\big[\,|0_{t}\rangle_\mathbf{q} \nonumber \\
&\!+\sum_{s,s'=1}^2\mathcal{C}_{\mathbf{q},ss'}(t)\hat{b}_{\mathbf{q},s}^{\dagger}(t)
\hat{d}_{-\mathbf{q},s'}^{\dagger}(t)|0_{t}\rangle_\mathbf{q}\\
&\!-\det \mathcal{C}_\mathbf{q}(t)\,
\hat{b}_{\mathbf{q},1}^{\dagger}(t)\hat{b}_{\mathbf{q},2}^{\dagger}(t)
\hat{d}_{-\mathbf{q},1}^{\dagger}(t)\hat{d}_{-\mathbf{q},2}^{\dagger}(t)
|0_{t}\rangle_\mathbf{q}\big]. \nonumber
\end{align}
Therefore, the probability amplitudes for zero pair production, one fermion-antifermion pair with spin indices $s,s'$, and two pair production are
\begin{align}
\mathcal{A}_\mathbf{q}^{(0)}&=\det\alpha_\mathbf{q}(t),\nonumber\\
\mathcal{A}_{\mathbf{q},ss'}^{(1)}&=-\det\alpha_\mathbf{q}(t)
\bigl[\alpha_\mathbf{q}^{-1}(t)\beta_\mathbf{q}(t)\bigr]^*_{ss'},
\label{eq:ProbAmpl2}\\
\mathcal{A}_\mathbf{q}^{(2)}&=-\det\alpha_\mathbf{q}(t)\det \bigl[\alpha_\mathbf{q}^{-1}(t)\beta_\mathbf{q}(t)\bigr]^*. \nonumber
\end{align}
The probability is
\begin{align}
\mathcal{P}_{\mathbf{q}}^{(0)}&=\bigl|\det\alpha_{\mathbf{q}}
\bigr|^{2}, \nonumber\\
\mathcal{P}_{\mathbf{q},ss'}^{(1)}&=|\det\alpha_{\mathbf{q}}|^{2}
\bigl|\bigl[\alpha_\mathbf{q}^{-1}(t)\beta_\mathbf{q}(t)\bigr]_{ss'}
\bigr|^{2}, \label{eq:Prob2} \\
\mathcal{P}_{\mathbf{q}}^{(2)}&=|\det\alpha_{\mathbf{q}}|^{2}
\bigl|\det\bigl[\alpha_\mathbf{q}^{-1}(t)\beta_\mathbf{q}(t)\bigr]
\bigr|^{2}=\bigl|\det\beta_{\mathbf{q}}\bigr|^2.
\nonumber
\end{align}
The unitary relations guarantee conservation of total probability:
\begin{align}
\mathcal{P}_{\mathbf{q}}^{(0)}+\sum_{s,s'=1}^2
\mathcal{P}_{\mathbf{q},ss'}^{(1)}+\mathcal{P}_{\mathbf{q}}^{(2)}=1.
\end{align}
The mean number of produced fermions in a single momentum mode $\mathbf{q}$ and spin $s$ is
\begin{equation}
\sum_{s'=1}^2\mathcal{P}_{\mathbf{q},ss'}^{(1)}
+\mathcal{P}_{\mathbf{q}}^{(2)}=1-[\alpha_\mathbf{q}(t)^\dagger
\alpha_\mathbf{q}(t)]_{ss},
\end{equation}
which is precisely the momentum distribution function given in Eq. (\ref{eq:MSF}).
The relative probability amplitude for produced pairs is defined as the probability amplitude for pair production divided by the vacuum-to-vacuum amplitude $\langle0_{t}|0_{t_0}\rangle_\mathbf{q}=\det \alpha_\mathbf{q}(t)$:
\begin{align}
\widetilde{\mathcal{A}}_\mathbf{q}^{\,(0)}&=1,\nonumber\\
\widetilde{\mathcal{A}}_{\mathbf{q},ss'}^{\,(1)}&=
-\bigl[\alpha_\mathbf{q}^{-1}(t)\beta_\mathbf{q}(t)\bigr]^*_{ss'},
\label{eq:RelaProbA2}\\
\widetilde{\mathcal{A}}_\mathbf{q}^{\,(2)}&=
-\det \bigl[\alpha_\mathbf{q}^{-1}(t)\beta_\mathbf{q}(t)\bigr]^*. \nonumber
\end{align}
The relative probability is
\begin{align}
\widetilde{\mathcal{P}}_{\mathbf{q}}^{\,(0)}&=1, \nonumber\\
\widetilde{\mathcal{P}}_{\mathbf{q},ss'}^{\,(1)}&=
\bigl|\bigl[\alpha_\mathbf{q}^{-1}(t)\beta_\mathbf{q}(t)\bigr]_{ss'}
\bigr|^{2}, \label{eq:RelaProb2} \\
\widetilde{\mathcal{P}}_{\mathbf{q}}^{\,(2)}&=
\bigl|\det\bigl[\alpha_\mathbf{q}^{-1}(t)\beta_\mathbf{q}(t)\bigr]
\bigr|^{2}.
\nonumber
\end{align}

From Eqs. (\ref{eq:DE})-(\ref{eq:DEInst}) and Eq. (\ref{eq:InitialCond2}), one can obtain the exact solutions $u_{\mathbf{q},s}(t)$ and $v_{-\mathbf{q},s}(t)$, and then, via Eqs. (\ref{eq:DEInst}) and (\ref{eq:Coeff2}), the Bogoliubov coefficients $\alpha_{\mathbf{q},ss'}(t)$ and $\beta_{\mathbf{q},ss'}(t)$.
Finally, using Eqs. (\ref{eq:MSF}), (\ref{eq:ProbAmpl2}), (\ref{eq:Prob2}), (\ref{eq:RelaProbA2}) and (\ref{eq:RelaProb2}), we can compute the corresponding momentum distribution, probability amplitudes and probabilities.

\section{Theoretical model and result analysis}
\label{sec:three}
The external field we considered is a spatially homogeneous and time-dependent electric field,
\begin{equation}\label{eq:EleField}
\mathbf{E}\left( t \right) =\frac{E_0}{\sqrt{1+\epsilon^2}}e^{-\frac{ t^2}{2\tau^2}}\left[ \cos \left( \Omega t \right) \mathbf{e}_x+\epsilon \sin \left( \Omega t \right) \mathbf{e}_y \right],
\end{equation}
where $E_0$ is the amplitude of the electric field for the linear polarization case ($\epsilon=0$), and the factor $1/\sqrt{1+\epsilon^2}$ is introduced to ensure the time-averaged intensity (or total energy) remains constant for varying ellipticity $\epsilon$.
The ellipticity $\epsilon$ ranges from $-1$ to $1$.
For $\epsilon=-1$ and $\epsilon=1$, the corresponding electric fields are right- and left-handed circularly polarized, respectively.
$\Omega$ is the field frequency.
$\tau$ denotes the pulse duration.
$\mathbf{e}_x$ and $\mathbf{e}_y$ are the unit vectors in the $x$ and $y$ directions, respectively.

In the following, we explore the vortex structures in the momentum distributions of multiple-pair production in linearly polarized, circularly polarized, and elliptically polarized electric fields.

\subsection{Linearly polarized electric fields}\label{sec:3a}

\begin{figure*}[!ht]
\centering
\includegraphics[width=\textwidth]{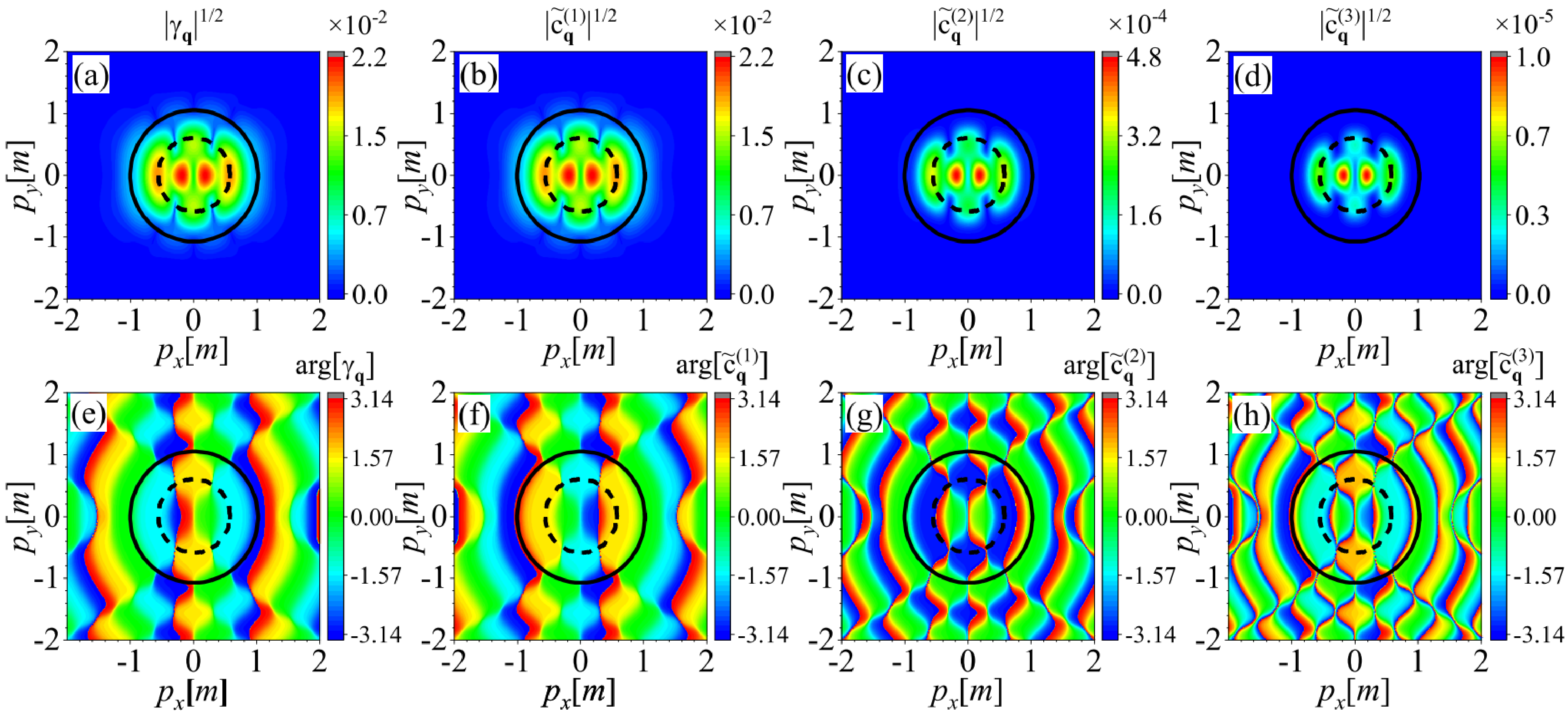}
\caption{The first row shows, from left to right, the momentum distributions for produced bosons ($\gamma_\mathbf{q}$) and for one, two, and three boson-antiboson pairs ($\widetilde{c}^{\,(1)}_\mathbf{q}$, $\widetilde{c}^{\,(2)}_\mathbf{q}$, and $\widetilde{c}^{\,(3)}_\mathbf{q}$, respectively), all produced in a linearly polarized electric field.
The second row shows the phase distributions of the corresponding probability amplitudes.
To clearly reveal the vortex structure, the momentum distribution function is displayed as its fourth root. The value of $q_z$ is set to $0.2m$.
The black lines in the figure indicate the positions of the multi-photon absorption rings.
The field parameters are $E_0=0.1E_\mathrm{cr}$, $\Omega=0.6m$, $\tau=10/m$, and $\epsilon=0$.
\label{B-W-Linearly}}
\end{figure*}

The momentum distributions (upper row) and the phase distributions of probability amplitudes (lower row) for boson-antiboson pairs produced in a weak linearly polarized electric field ($E_0=0.1E_{\mathrm{cr}}$) are presented in Fig. \ref{B-W-Linearly}. The upper row (Figs. \ref{B-W-Linearly}(a)-\ref{B-W-Linearly}(d)) exhibits a regular pattern of concentric rings, see the black lines. Figure \ref{B-W-Linearly}(a) shows the momentum distribution of the mean number of produced bosons ($\gamma_\mathbf{q}$), while Figs. \ref{B-W-Linearly}(b), \ref{B-W-Linearly}(c), and \ref{B-W-Linearly}(d) correspond to the momentum distributions of one, two, and three produced boson-antiboson pairs ($\widetilde{c}^{\,(1)}_\mathbf{q}$, $\widetilde{c}^{\,(2)}_\mathbf{q}$, and $\widetilde{c}^{\,(3)}_\mathbf{q}$, respectively). These well-defined concentric rings are multi-photon absorption rings. According to energy conservation, the total energy required to create one pair must correspond to the absorption of an integer number of photons, i.e., $2\sqrt{\mathbf{p}^2+m_*^2}=N\Omega$, where $m_*$ is the effective mass and $N$ is the number of absorbed photons. Here the field frequency $\Omega=0.6m$. Thus, in momentum space, the innermost ring corresponds to the absorption of four photons, and each successive outer ring corresponds to one additional photon (i.e., five-photon processes).
Inside the four-photon absorption ring, a smaller ring is visible at a radius smaller than that of the four-photon ring. It seems to correspond to a three-photon absorption process. According to energy conservation, however, $3\Omega=1.8m<2m$, so a strict three-photon process is forbidden.
This apparent contradiction arises from the finite pulse duration.
The Gaussian pulse has $\tau=10/m$, which gives a spectral width $\Delta\Omega\sim1/\tau=0.1m$. The bandwidth of the three-frequency sum is estimated as $\sqrt{3}\Delta\Omega\simeq0.17m$, while the threshold deficit is $2m-3\Omega=0.2m$. Therefore the high-frequency tail of the pulse can activate a weak sub-threshold three-photon-like ring. Note that this suppression is relative to an allowed three-photon process and does not imply that the three-photon-like channel must be weaker than the four-photon process. The ring should weaken or disappear as $\tau$ increases or in the monochromatic limit. For $\Omega>2m/3\simeq0.667m$, the three-photon process becomes genuinely allowed.
For multi-pair production, the total energy equation for producing $n$ pairs is $n$ times that for a single pair, and the number of absorbed photons is likewise multiplied by $n$. However, dividing both sides of the energy equation by $n$ yields the same relation between momentum and photon number as in the single-pair case. This is precisely why the positions of the multi-photon rings in Figs. \ref{B-W-Linearly}(c) and \ref{B-W-Linearly}(d) (two- and three-pair production) coincide exactly with those in Fig. \ref{B-W-Linearly}(b) (single-pair production).

On these multi-photon rings, nodal line structures are clearly distributed. From the phase distributions of the boson pair-production probability amplitude shown in Figs. \ref{B-W-Linearly}(e)-\ref{B-W-Linearly}(h), one can see that there is no phase winding along these nodal lines, so they are not vortices. Their physical origin is the interference between pair-creation paths at different times within one period of the oscillating electric field. When these paths are superimposed in momentum space, they cancel due to a phase difference of $\pi$, thereby forming nodal lines. This phenomenon is highly analogous in physical mechanism to the interference nodes in strong-field ionization in atomic and molecular physics. A further comparison of Figs. \ref{B-W-Linearly}(a) and \ref{B-W-Linearly}(b) shows that the two are not only morphologically identical but also nearly equal in magnitude. This arises because, in the weak-field regime, the modulus $|\beta_\mathbf{q}/\alpha_\mathbf{q}|$ of the Bogoliubov-coefficient ratio is much smaller than unity, so that the probabilities of higher-order multi-pair production, which are proportional to $|\beta_\mathbf{q}/\alpha_\mathbf{q}|^{2n}$, are exponentially strongly suppressed. The contribution of multi-pair production is therefore negligible, and the momentum distribution of the mean particle number is essentially dominated by single-pair production.

\begin{figure*}[!ht]
\centering
\includegraphics[width=\textwidth]{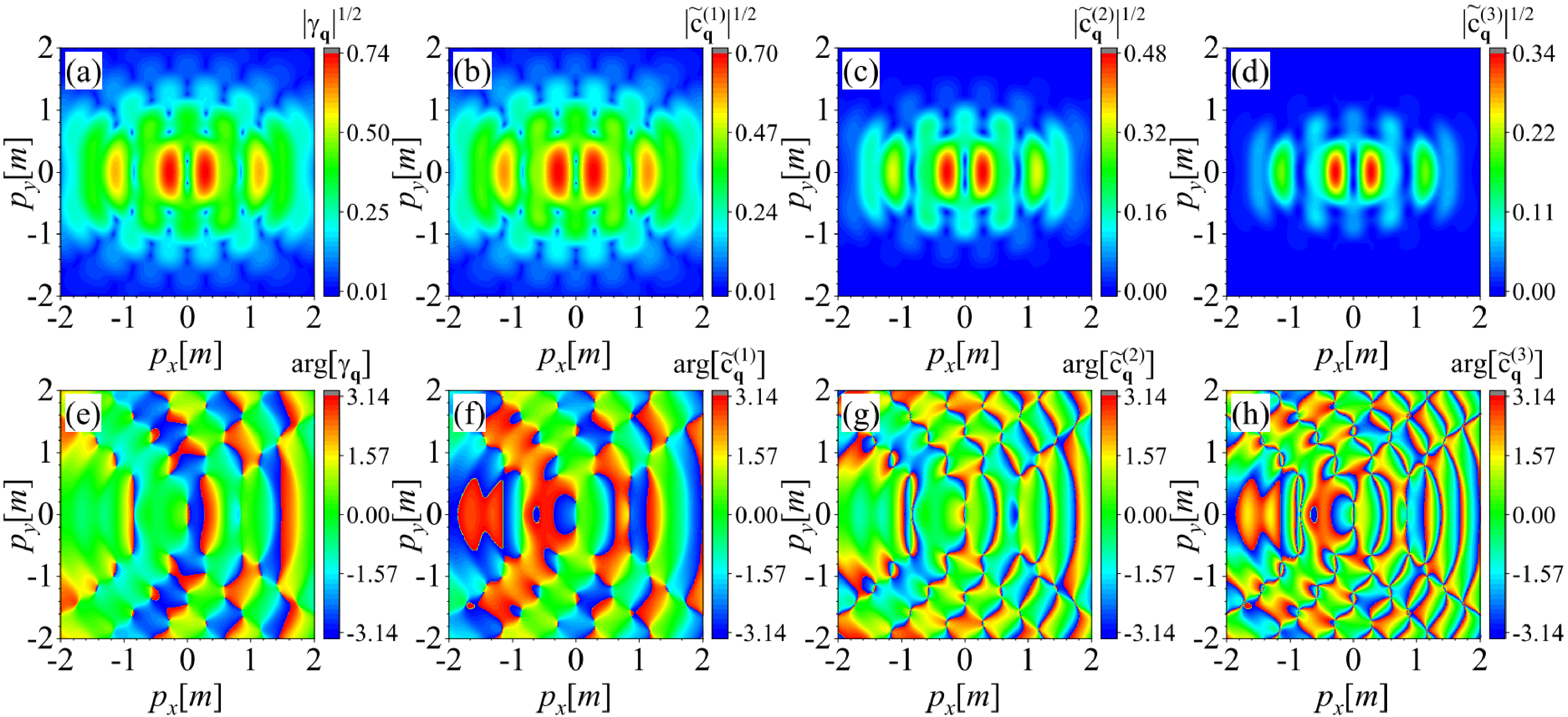}
\caption{Same as Fig. \ref{B-W-Linearly}, but for $E_0=1E_\mathrm{cr}$.
\label{B-S-Linearly}}
\end{figure*}

For a strong linearly polarized field ($E_0=1E_{\mathrm{cr}}$), a regime in which multi-photon absorption and tunneling effects coexist, the momentum distributions of created boson-antiboson pairs shown in Figs. \ref{B-S-Linearly}(a)-\ref{B-S-Linearly}(d) change dramatically: the originally discrete multi-photon rings are broadened and blurred, evolving into highly dense interference fringes, reflecting the significantly enhanced contribution of the continuous energy spectrum due to tunneling. At the same time, the originally continuous nodal lines become unstable and break up into isolated nodal points. From the phase distributions of the pair-production probability amplitude in Figs. \ref{B-S-Linearly}(e)-\ref{B-S-Linearly}(h), one can see that these isolated nodal points appear in pairs topologically, and the phase accumulates $2\pi$ around each point, indicating that they correspond unambiguously to vortex-antivortex pairs with topological charges $+1$ and $-1$. Comparing Fig. \ref{B-S-Linearly}(a) (momentum distribution of the mean particle number) with Fig. \ref{B-S-Linearly}(b) (single-pair production), one finds that the overall interference skeleton remains similar, but the values in Fig. \ref{B-S-Linearly}(a) are noticeably larger than those in Fig. \ref{B-S-Linearly}(b). This is because, in the strong-field regime, multi-pair production is no longer strongly suppressed, and multi-pair processes contribute substantially more to the average, thereby widening the numerical gap between the two distributions. Moreover, the values in Figs. \ref{B-S-Linearly}(c) and \ref{B-S-Linearly}(d) do not drop by orders of magnitude relative to Fig. \ref{B-S-Linearly}(b), further confirming the significant enhancement of multi-pair production probability in strong fields.

A further comparison of the vortex structures in single-pair (Fig. \ref{B-S-Linearly}(b)) and multi-pair (Figs. \ref{B-S-Linearly}(c) and \ref{B-S-Linearly}(d)) production in Fig. \ref{B-S-Linearly} reveals directly that the vortex topological charge for two-pair production is twice that for single-pair production, and for three-pair production it is three times as large. This feature originates from the complex phase property of the relative probability amplitude $\widetilde{c}^{\,(n)}_\mathbf{q} = (\beta^*_\mathbf{q}/\alpha^*_\mathbf{q})^n$. Because the probability amplitude for $n$-pair production is the $n$-th power of the single-pair amplitude, encircling a vortex singularity in momentum space produces a phase winding of $2\pi$ for single-pair production (topological charge 1), whereas for $n$-pair production the phase winding is multiplied by $n$ (topological charge $n$).
%Physically, this means that in the strong-field regime, the total angular-momentum structure carried by multi-pair production in momentum space undergoes an integer-fold enhancement.

%Finally, a combined comparison of Figs. \ref{B-W-Linearly} and \ref{B-S-Linearly} clearly shows that, as the field strength increases, the system undergoes a notable topological phase transition. In the weak-field case of Fig. \ref{B-W-Linearly}, destructive interference appears in the form of continuous nodal lines; in the strong-field case of Fig. \ref{B-S-Linearly}, these nodal lines break and evolve into isolated vortex pairs carrying positive and negative topological charges. This evolution reveals a transition in vacuum decay from pure energy-shell interference to the generation of topological defects with angular-momentum character, providing a new perspective for studying the nonperturbative dynamics of the vacuum structure in strong-field quantum electrodynamics.

\begin{figure*}[!ht]
\centering
\includegraphics[width=\textwidth]{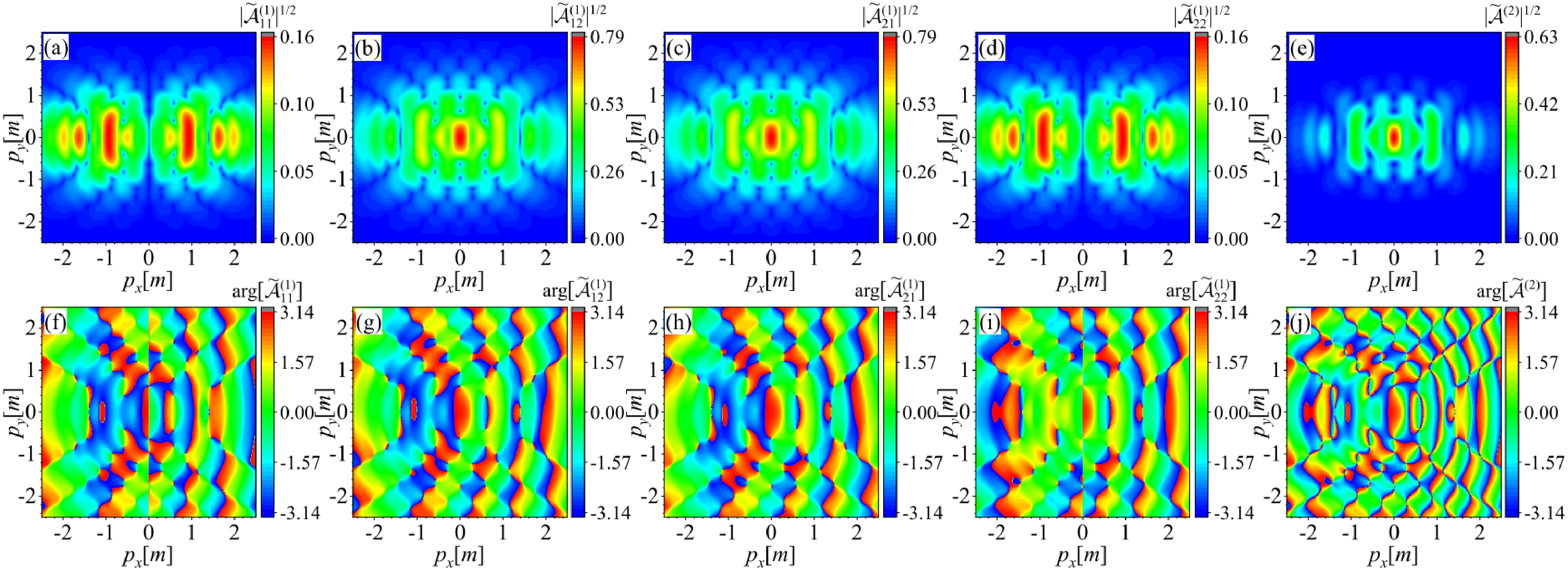}
\caption{The first row shows, from left to right, the four spin-resolved momentum distributions for one fermion-antifermion pair ($\widetilde{\mathcal{A}}^{(1)}_{\mathbf{q},ss'}$) and the momentum distribution for two fermion-antifermion pairs ($\widetilde{\mathcal{A}}^{(2)}_\mathbf{q}$), both produced in a linearly polarized electric field.
The second row shows the phase distributions of the corresponding probability amplitudes.
Panels (a) and (f) correspond to $s=1, s'=1$; panels (b) and (g) to $s=1, s'=2$; panels (c) and (h) to $s=2, s'=1$; panels (d) and (i) to $s=2, s'=2$; and panels (e) and (j) to the two-pair case.
To clearly reveal the vortex structure, the momentum distribution function is displayed as its fourth root.
The value of $q_z$ is set to $0.2m$.
The field parameters are $E_0=1E_\mathrm{cr}$, $\Omega=0.6m$, $\tau=10/m$, and $\epsilon=0$.
\label{F-S-Linearly}}
\end{figure*}

Since the pair production process for fermions in a weak linearly polarized electric field (e.g., $E_0=0.1E_{\mathrm{cr}}$) is similar to the bosonic case --- single-pair production dominates overwhelmingly, multi-pair production is strongly suppressed, and only nodal line structures rather than vortex structures appear --- we do not present the results for fermion pair production in the weak-field regime. In addition, the single-pair momentum distribution $|\gamma_{\mathbf{q}, s's}|^2$ calculated from the Dirac sea concept is almost identical to the relative probability amplitude $\widetilde{\mathcal{A}}^{(1)}_{\mathbf{q}, ss'}$ for single-pair production in the weak-field regime. In the strong-field regime, although the two differ in magnitude, the differences in their structures (such as the positions and characteristics of vortices) remain very small. Therefore, we also do not show the single-pair momentum distribution $|\gamma_{\mathbf{q}, s's}|^2$ obtained from the Dirac sea concept. To display more clearly the vortex structures in multi-pair fermion-antifermion production and to compare with the bosonic case, Fig. \ref{F-S-Linearly} presents only the momentum distributions and phase distributions of the relative probability amplitude in a strong linearly polarized field.

From the first row of Fig. \ref{F-S-Linearly}, one can see that the momentum distribution for single fermion pair production exhibits four spin combinations. All of them are symmetric about $p_y=0$, since the external electric field is axially symmetric.
Furthermore, the two distributions with identical spin indices (i.e., $s=1, s'=1$ in Fig. \ref{F-S-Linearly}(a) and $s=2,s'=2$ in Fig. \ref{F-S-Linearly}(d)) are completely identical, and the two distributions with different spin indices (i.e., $s=1,s'=2$ in Fig. \ref{F-S-Linearly}(b) and $s=2,s'=1$ in Fig. \ref{F-S-Linearly}(c)) are also completely identical.
This follows from the tracelessness of the relative amplitude matrix $\widetilde{A}^{(1)}_{\mathbf q, ss'}$ for single-pair production  and the reflection symmetry with respect to the $x$-$z$ plane.
At the same time, one notices that the peak values of the momentum distributions in Figs. \ref{F-S-Linearly}(b) and \ref{F-S-Linearly}(c) (about 0.79) are higher than those in Figs. \ref{F-S-Linearly}(a) and \ref{F-S-Linearly}(d) (about 0.16), indicating that pair production channels corresponding to different spin-index combinations have a higher production probability in the strong-field regime. This is consistent with the larger squared matrix elements for these combinations in Eq. (\ref{eq:RelaProb2}). It can also be understood through the spin selection rules obtained from the relative amplitude matrix $\widetilde{A}^{(1)}_{\mathbf q, ss'}$: The momentum distributions for the two single-pair production processes with identical spin indices exhibit a nodal surface at $q_z=0$, where they vanishes. Although the distributions no longer vanish when $q_z\neq 0$, they remain suppressed, and the closer $q_z$ is to zero, the stronger the suppression. Finally, comparing the peak values of the momentum distributions for two-pair and single-pair production shows that multi-pair production is much less suppressed in the strong-field regime and can no longer be neglected.

A horizontal comparison between the momentum distributions in Fig. \ref{F-S-Linearly} and those in Fig. \ref{B-S-Linearly} (bosonic case) directly reveals different quantum statistical properties of the two systems. Comparing the fermion momentum distributions for different spin-index combinations (Figs. \ref{F-S-Linearly}(b) and \ref{F-S-Linearly}(c)) with the bosonic single-pair distribution, one finds that the positions of the maxima in the fermion distributions correspond to the positions of the minima in the bosonic distribution, and vice versa, giving rise to a clear ``bright-dark interchange'' phenomenon. This is not limited to single-pair production. The same inverse relation between maxima and minima also holds in the two-pair momentum distributions (Fig. \ref{F-S-Linearly}(e) versus Fig. \ref{F-S-Linearly}(d)). This significant difference originates from the fact that fermions obey Fermi-Dirac statistics (with antisymmetric wave functions), whereas bosons obey Bose-Einstein statistics (with symmetric wave functions).
%The Pauli exclusion principle for fermions tends to make them repel each other and avoid occupying the same state, whereas bosons tend to cluster together, directly leading to opposite interference patterns when they are coherently superimposed in momentum space.

An analysis of the phase distributions in the second row (Figs. \ref{F-S-Linearly}(f)-\ref{F-S-Linearly}(i)) further reveals the imprint of spin degrees of freedom on the phase structure. In the phase maps for single-pair production with identical spin indices (Figs. \ref{F-S-Linearly}(f) and \ref{F-S-Linearly}(i)), both of them are symmetric about $p_y=0$, present a clear nodal surface structure at $p_x=0$, and differ in phase $\pi$.
This results reflect the symmetry constraints on the wave function for specific spin combinations and the spin selection rules in fermion pair production.
For different spin-index combinations (Figs. \ref{F-S-Linearly}(g) and \ref{F-S-Linearly}(h)), neither of the two phase maps is symmetric about $p_y=0$ or shows a nodal surface, but they are mirror images of each other with respect to $p_y=0$, and this mirror symmetry arises from the reflection symmetry with respect to the $x$-$z$ plane.
Despite the spin-specific nodal surfaces, the momentum distributions and phase distributions of fermion pair production still exhibit clear and complex vortex structures similar to those in the bosonic case. This indicates that vortex generation is a universal topological property of strong-field vacuum decay, independent of particle statistics, but the fermionic system possesses richer internal structure owing to the introduction of spin indices. It is worth emphasizing that, because fermions are subject to the Pauli exclusion principle, at most two pairs can be produced in a single momentum mode (corresponding to $\widetilde{\mathcal{P}}_{\mathbf{q}}^{(2)}$ in the formula), so the topological structure of fermion multi-pair production in phase space does not exhibit higher-order topological charges as in the bosonic case, but it still displays rich strong-field interference features.

\subsection{Circularly polarized electric fields}\label{sec:3b}

\begin{figure*}[!ht]
\centering
\includegraphics[width=\textwidth]{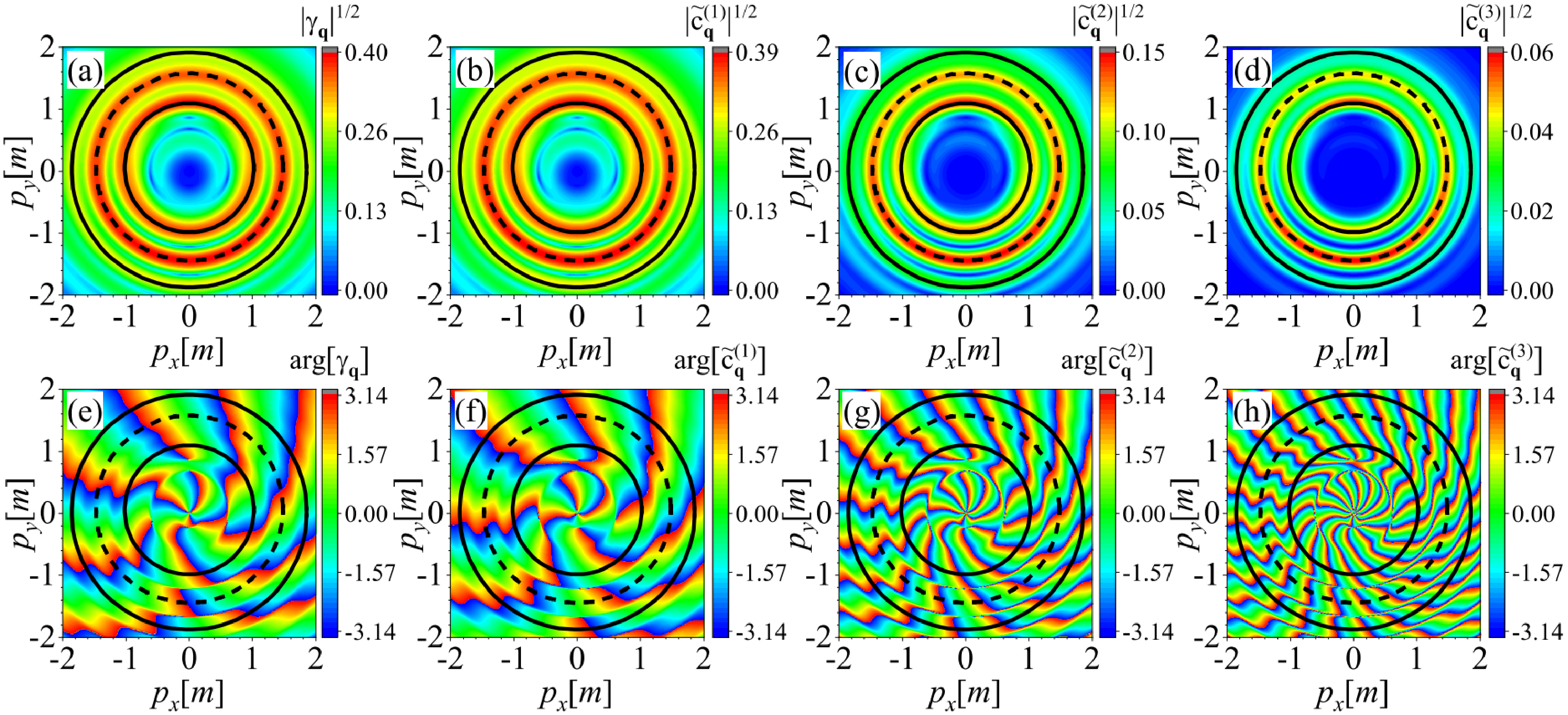}
\caption{Same as Fig. \ref{B-S-Linearly}, but for a circularly polarized electric field, i.e., $\epsilon=1$. The black lines in the figure indicate the positions of the multi-photon absorption rings.
\label{B-S-Circularly}}
\end{figure*}

In a strong circularly polarized electric field, we first analyze the boson pair production shown in Fig. \ref{B-S-Circularly}. In the momentum distributions, because of the rotational symmetry of the circularly polarized field, the interference nodal lines that appeared in the linearly polarized case are completely smoothed out, and are replaced by extremely clear and continuous concentric multi-photon rings. The black lines in the figure indicate the ring positions, and a distinct dark core appears at the center. From the phase distributions, one finds a stable central vortex at the origin, and as the momentum crosses successive multi-photon rings outward, the phase winding number increases by one each time for Fig. \ref{B-S-Circularly}(a), with the corresponding topological charge also increasing by one. This is because in a left-handed circularly polarized field each photon carries not only energy but also an angular momentum of $+1$; according to angular momentum conservation, the orbital angular momentum $z$ component $L_z$ of the produced boson pair must equal the total number of absorbed photons $N$. Therefore, from the innermost ring outward, the number of absorbed photons $N$ increases successively, and the topological charge is directly equal to the number of absorbed photons. For multi-pair production, although the peak values in Figs. \ref{B-S-Circularly}(c) and \ref{B-S-Circularly}(d) are smaller than that of single-pair production in Fig. \ref{B-S-Circularly}(b), the suppression is significantly weakened in the strong-field regime, and the difference is much smaller than in the weak-field case. The positions of the multi-photon rings coincide exactly with those of single-pair production, because the particle-number factors cancel in the energy conservation equation, so the ring positions are determined solely by energy conservation and are independent of the particle number. The phase distributions show that when crossing the same multi-photon ring, the topological charge for two-pair production is exactly twice that for single-pair production, and that for three-pair production is three times as large, consistent with the physical mechanism whereby the power-law relation of the probability amplitude determines the multiple of the phase winding.

\begin{figure*}[!ht]
\centering
\includegraphics[width=\textwidth]{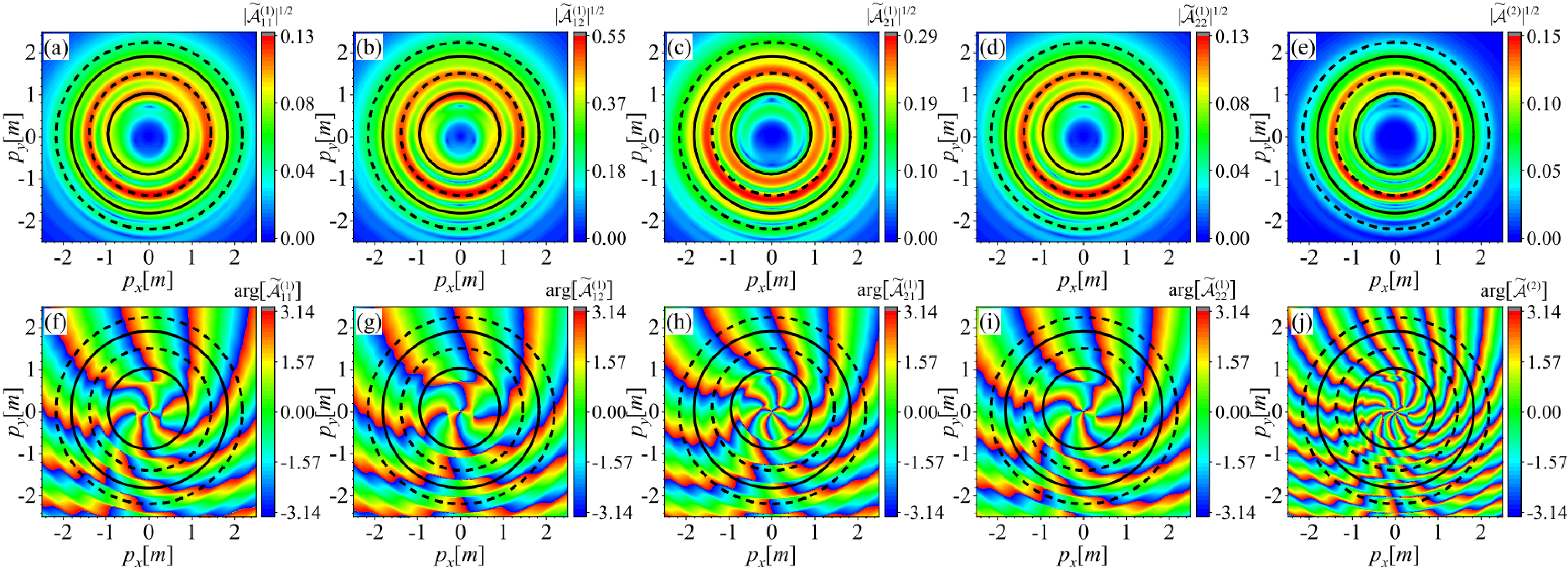}
\caption{Same as Fig. \ref{F-S-Linearly}, but for a circularly polarized electric field, i.e., $\epsilon=1$. The black lines in the figure indicate the positions of the multi-photon absorption rings.
\label{F-S-Circularly}}
\end{figure*}

The results for fermion pair production are shown in Fig. \ref{F-S-Circularly}.
%It should first be noted that the positions of the multi-photon rings for fermions are exactly the same as those for bosons, because the ring positions are determined solely by energy conservation (i.e., $2\sqrt{m^2+p^2}=N\Omega$) and are independent of the quantum statistics obeyed by the particles.
In the momentum distributions, the four spin combinations for single-pair production exhibit complex features: the combinations with identical spin indices (Figs. \ref{F-S-Circularly}(a) and \ref{F-S-Circularly}(d)) are completely identical, whereas the combinations with different spin indices (Figs. \ref{F-S-Circularly}(b) and \ref{F-S-Circularly}(c)) show significant morphological differences. A precise comparison shows that the peak value in Fig. \ref{F-S-Circularly}(b) is larger than that in Fig. \ref{F-S-Circularly}(c), with Figs. \ref{F-S-Circularly}(a) and \ref{F-S-Circularly}(d) lying in between; at the same time, the radii of the multi-photon rings in Fig. \ref{F-S-Circularly}(b) are slightly smaller than those in Fig. \ref{F-S-Circularly}(c), and Figs. \ref{F-S-Circularly}(a) and \ref{F-S-Circularly}(d) are again intermediate. From the phase distributions, if one examines the same order of multi-photon ring, the orbital angular momentum $z$ component $L_z$ in Fig. \ref{F-S-Circularly}(b) is larger by 1 than that in Figs. \ref{F-S-Circularly}(a) and \ref{F-S-Circularly}(d), and those in Figs. \ref{F-S-Circularly}(a) and \ref{F-S-Circularly}(d) are larger by 1 than that in Fig. \ref{F-S-Circularly}(c). This ordering can be explained rigorously by angular momentum conservation: the total angular momentum of a fermion pair includes both orbital and spin parts. For processes absorbing the same number of photons $N$, total angular momentum conservation requires $L_z + S_z = N$. Figure \ref{F-S-Circularly}(b) corresponds to the production of fermion pairs with total spin $z$ component $S_z=1$, for which $L_z = N-1$; Figs. \ref{F-S-Circularly}(a) and \ref{F-S-Circularly}(d) correspond to $S_z=0$, for which $L_z=N$; and Fig. \ref{F-S-Circularly}(c) corresponds to $S_z=-1$, for which $L_z = N+1$.
Note that, in the probability amplitude for the production of a single fermion-antifermion pair, the spin index assigned to the antifermion does not label its spin projection along the $z$ axis. That projection is instead labeled by the complementary spin index. For instance, if the antifermion is assigned the spin index $s=1$, its spin projection along the $z$ axis is the one labeled by $s=2$, and conversely \cite{Jiang20262}.

From this, two physical phenomena can be discussed in depth. The first is the regularity of the ring radii: the radii of the multi-photon rings are positively correlated with the orbital angular momentum $L_z$, because a larger orbital angular momentum implies a higher centrifugal barrier for the particles in momentum space, so the radius required to satisfy energy conservation also increases. Therefore, Fig. \ref{F-S-Circularly}(c) has the largest radii, Fig. \ref{F-S-Circularly}(b) the smallest, and Figs. \ref{F-S-Circularly}(a) and \ref{F-S-Circularly}(d) are intermediate. The second is the counterintuitive behavior of the production probabilities: if one relies only on the intuitive expectation that a larger orbital angular momentum makes production more difficult, Fig. \ref{F-S-Circularly}(c) (with the largest $L_z$) should have the smallest peak, but in the actual observations the peaks in Figs. \ref{F-S-Circularly}(a) and \ref{F-S-Circularly}(d) are the smallest. This indicates that a more dominant selection rule is at work: the suppression due to the spin selection rules for fermion pair production discussed earlier in the linearly polarized field. Specifically, at $q_z=0$, the production of fermion pairs with total spin $S_z=0$ is strictly forbidden; in the actual calculation with $q_z=0.2m\neq0$, this process is strongly suppressed, and the suppression becomes more pronounced as $q_z$ approaches zero. It is particularly important to note that this suppression is global and there is no nodal surface structure at $p_x=0$. It is precisely the overall suppression of this spin channel that leads to the anomalously small peaks in Figs. \ref{F-S-Circularly}(a) and \ref{F-S-Circularly}(d).

Comparison of the boson and fermion results in Figs. \ref{B-S-Circularly} and \ref{F-S-Circularly} shows that their topological structures agree only for $S_z=0$. The multi-photon ring positions are determined not solely by energy conservation, but also modulated by orbital angular momentum. For a single fermion-antifermion pair, among the four spin combinations, only the $S_z=0$ cases reproduce the bosonic ring positions, with no spatial shift or inversion of maxima and minima. The $S_z=+1$ rings have smaller radii than the bosonic rings, while the $S_z=-1$ rings have larger radii. Likewise, the topological charge of the central vortex equals the number of absorbed photons only for $S_z=0$, where it parallels the bosonic result. For $S_z=\pm1$, this equality does not hold. Multi-pair production remains unsuppressed in the strong-field regime for both bosons and fermions. Quantum statistical effects do not shift the momentum distributions spatially. They appear instead in spin-dependent features. In particular, the fermion momentum peaks depend more strongly on the spin channel than the bosonic peaks, as seen in the pronounced suppression of a specific $S_z=0$ channel.
Under a weak circularly polarized electric field, the momentum distributions of the created particle-antiparticle pairs also present a central vortex structure, and results analogous to those above are obtained. The only difference is that the momentum peaks for multi-pair production are much smaller than those for single-pair production.

\subsection{Elliptically polarized electric fields}\label{sec:3c}

\begin{figure*}[!ht]
\centering
\includegraphics[width=\textwidth]{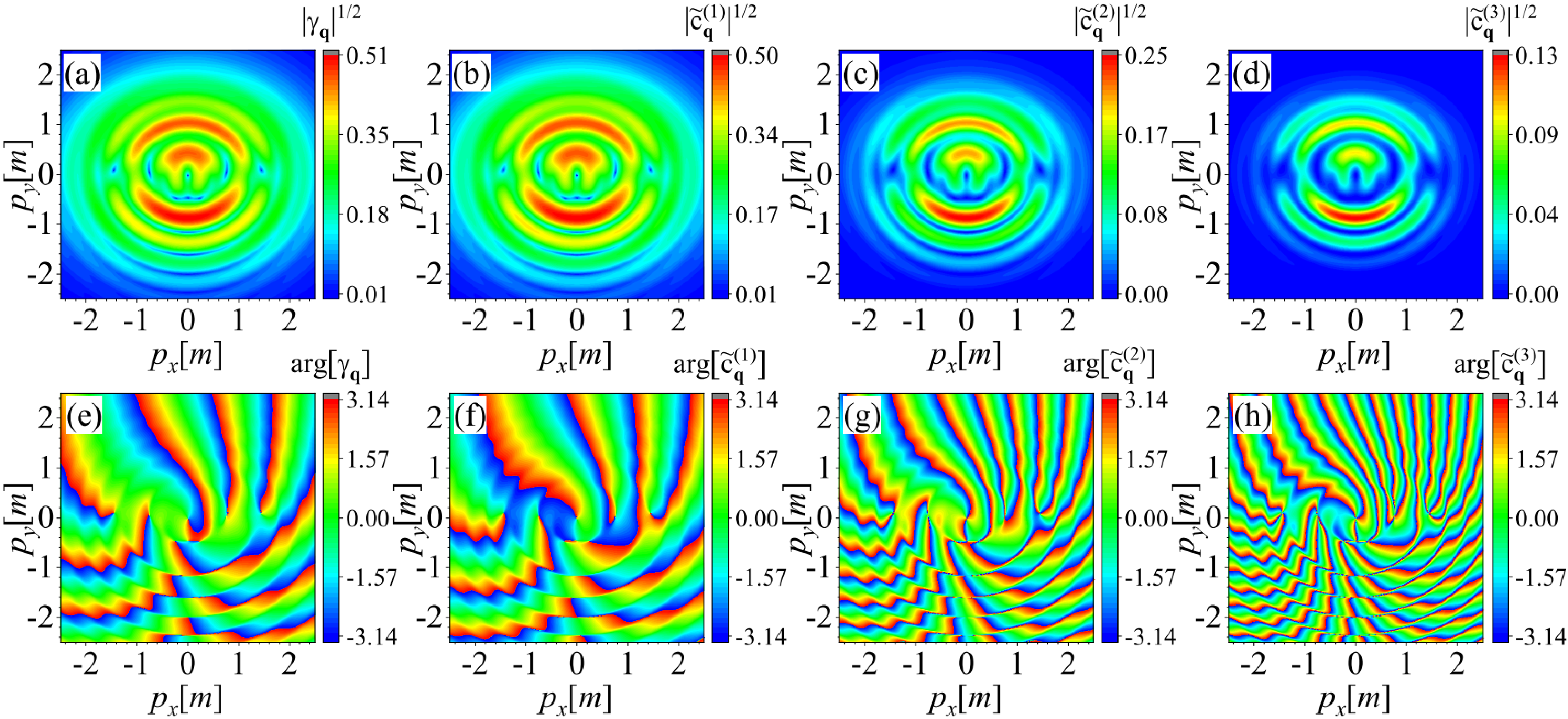}
\caption{Same as Fig. \ref{B-S-Linearly}, but for a elliptically polarized electric field, i.e., $\epsilon=0.5$.
\label{B-S-Elliptically}}
\end{figure*}

We first analyze the case of boson pair production shown in Fig. \ref{B-S-Elliptically}.
In the momentum distributions, since the ellipticity $\epsilon=0.5$ lies between the linear and circular polarization limits, the electric field amplitudes in the $x$ and $y$ directions differ, thereby breaking the perfect rotational symmetry of the circularly polarized case.
As a result, although the multi-photon rings remain clearly visible in Figs. \ref{B-S-Elliptically}(a)-\ref{B-S-Elliptically}(d), their shapes are no longer perfectly circular but exhibit a slight scalloped or elliptically stretched distortion.
It should be emphasized that the positions of these multi-photon rings are basically determined by the energy conservation equation, so the ring positions for multi-pair production (Figs. \ref{B-S-Elliptically}(c) and \ref{B-S-Elliptically}(d)) basically coincide with those for single-pair production (Fig. \ref{B-S-Elliptically}(b)).
In the strong-field regime, the peak values for multi-pair production (0.25 and 0.13) are lower than that for single-pair production (0.50), but the suppression is not pronounced.
From the phase distributions in Figs. \ref{B-S-Elliptically}(e)-\ref{B-S-Elliptically}(h), a stable principal vortex exists at the center.
At the same time, due to the strong field and the directional nature of the elliptical polarization, discrete vortex points appear in the outer region along specific directions.
Comparing the phase distributions for one-, two-, and three-pair production, one finds that the vortex topological charges for boson multi-pair production are precisely integer multiples (twofold and threefold) of that for single-pair production.
This originates from the fact that the relative probability amplitude for bosons strictly satisfies the power-law relation $(\beta^*_\mathbf{q}/\alpha^*_\mathbf{q})^n$, so the number of phase windings increases linearly with $n$.

\begin{figure*}[!ht]
\centering
\includegraphics[width=\textwidth]{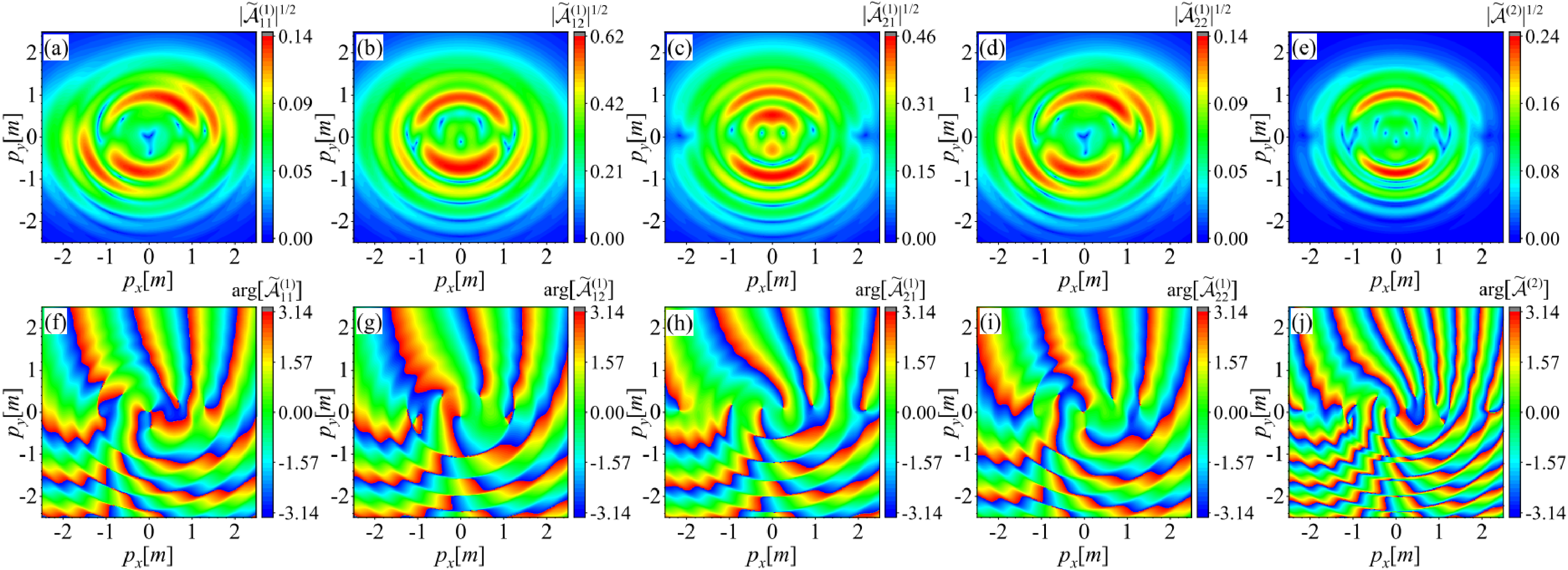}
\caption{Same as Fig. \ref{F-S-Linearly}, but for a elliptically polarized electric field, i.e., $\epsilon=0.5$.
\label{F-S-Elliptically}}
\end{figure*}

The case of fermion pair production is shown in Fig. \ref{F-S-Elliptically}.
The positions of the multi-photon rings are likewise essentially determined by energy conservation, and the ring positions in Fig. \ref{F-S-Elliptically}(e) for two-pair production basically coincide with those for single-pair production (Figs. \ref{F-S-Elliptically}(b), \ref{F-S-Elliptically}(c), and \ref{F-S-Elliptically}(d)).
In the momentum distributions, the identical-spin combinations (Figs. \ref{F-S-Elliptically}(a) and \ref{F-S-Elliptically}(d)) are exactly the same, while the different-spin combinations (Figs. \ref{F-S-Elliptically}(b) and \ref{F-S-Elliptically}(c)) show clear differences in intensity, which again arises from angular momentum conservation and the spin selection rules.
In the phase distributions (Figs. \ref{F-S-Elliptically}(f)-\ref{F-S-Elliptically}(j)), the vortex positions for different spin combinations exhibit notable differences, because after the elliptical polarization breaks the rotational symmetry, the dynamical trajectories of different spin channels diverge.
However, comparing the phase maps for single-pair production (Fig. \ref{F-S-Elliptically}(b)) and two-pair production (Fig. \ref{F-S-Elliptically}(e)), one finds that the vortex topological charge for two-pair production does not appear as twice that for single-pair production. The fundamental reason is that the probability amplitude for two-pair fermion-antifermion production is determined by the determinant of a matrix, $\det[\widetilde{A}^{(1)}_{\mathbf q, ss'}]$ (Eq. (56)).
Under strong linearly polarized fields (Figs. \ref{B-S-Linearly} and \ref{F-S-Linearly} and strong circularly polarized fields (Figs. \ref{B-S-Circularly} and \ref{F-S-Circularly}, the high symmetry ensures that the vortices of the matrix elements $\widetilde{\mathcal{A}}^{(1)}_{11}\widetilde{\mathcal{A}}^{(1)}_{22}$ and $\widetilde{\mathcal{A}}^{(1)}_{12}\widetilde{\mathcal{A}}^{(1)}_{21}$ either coincide in momentum space or have strictly correlated phases, so that evaluating the determinant is equivalent to a precise superposition of phases, which stably preserves the factor-of-two relation.
However, in an elliptically polarized field, the asymmetry between the field strengths in the $x$ and $y$ directions causes the vortices of different spin channels to be clearly displaced in momentum space.
In this case, the phase distributions of the two terms in the determinant are no longer aligned, and their subtraction induces strong destructive interference and phase reconstruction, destroying the original simple superposition law and making the resulting vortex topological structure no longer an integer multiple of the single-pair case.
This vividly illustrates the influence of symmetry breaking on topological phase transitions in fermionic many-body systems.

Finally, comparing Figs. \ref{B-S-Elliptically} and \ref{F-S-Elliptically} leads to a key conclusion: the elliptically polarized field, as an intermediate state between linear and circular polarization, can possess topological features of both cases for a suitable ellipticity, while also displaying behavior unique to elliptical polarization.
In the bosonic system, although the shapes of the multi-photon rings are distorted, the strict power-law structure of the probability amplitude ensures that the topological charge inheritance (integer multiples) in multi-pair production remains robust.
In the fermionic system, however, owing to the Pauli exclusion principle and the presence of spin degrees of freedom, the topological charges of fermion multi-pair production are strongly modulated by the interference between different terms in the matrix determinant.
The variation of the ellipticity essentially controls the degree of separation between the vortices of different spin channels, thereby providing an excellent tunable platform for investigating in depth the complex interference effects and topological evolution in fermion multi-pair production under strong fields.

\section{Conclusion and discussion}
\label{sec:four}

Using the Bogoliubov transformation, we have studied the vortex structures in the momentum distributions of multiple boson-antiboson and fermion-antifermion pairs produced from vacuum by a spatially homogeneous and time-dependent electric field with arbitrary polarization.

For linear polarization in the weak-field limit, multi-photon absorption dominates for both bosons and fermions and favors single-pair production.
Both single- and multi-pair momentum distributions exhibit multi-photon rings and nodal lines within the rings.
For single-pair fermion production, spin selection rules impose nodal surfaces.
Away from these surfaces, the maxima and minima of the bosonic and fermionic distributions are interchanged, reflecting Bose enhancement versus Pauli blocking.
In multi-pair production, such spin-selection nodal surfaces are absent, and the maxima and minima are directly interchanged. The bosonic amplitude is suppressed by a power law, whereas the fermionic amplitude has a determinant structure.
As the field approaches $E_{\mathrm{cr}}$, tunneling and multi-photon processes coexist. The continuous nodal lines split into vortex-antivortex pairs in both single- and multi-pair distributions, and the suppression of multi-pair production weakens.
After this splitting, the topological charge of the vortices in the multi-pair distribution is an integer multiple of the corresponding single-pair value for both bosons and fermions.

Under circular polarization, the multi-photon rings are smooth and exhibit no nodal lines, and a central vortex is present.
The spin-selection-induced nodal surfaces that appear in fermion-pair creation under linear polarization are also absent.
For single-pair production, the topological charge of the central vortex increases by one per absorbed photon, ring by ring.
For fermions, spin enters via $L_z+S_z=N$, where $N$ is the absorbed-photon number, $L_z$ is the orbital angular momentum projection of the pair along the propagation direction, and $S_z$ is the total spin projection of the pair.
The $S_z=+1$ channel has smaller orbital angular momentum and a lower centrifugal barrier. Its peak is higher than that in the bosonic case, and its ring radii are the smallest.
The $S_z=-1$ channel has the highest barrier and a weaker peak, whereas the $S_z=0$ channel is suppressed by spin selection rules.
For multi-pair production, the central vortex persists.
For a given multi-photon ring, the bosonic topological charge is an integer multiple of the single-pair topological charge, and the fermionic topological charge is an integer multiple of that of the $S_z=0$ channel.

For elliptical polarization, rotational symmetry is only partially broken.
The positions of the multi-photon rings are basically fixed by energy conservation and are independent of the number of pairs, but the rings are distorted, and a central vortex coexists with discrete vortices. For bosons, the symmetric power structure preserves integer scaling of the vortex charge in multi-pair production.
For fermions, spin-channel and orbital-angular-momentum dynamics displace the single-pair vortices in momentum space.
The two-pair production amplitude is a determinant; its terms undergo phase reconstruction and interference that can enhance or suppress individual contributions.
Consequently, the multi-pair topological charge is no longer a simple integer multiple of the single-pair charge.
This shows how symmetry breaking, spin, and many-body interference modify the topological structure.

In summary, varying the polarization provides a way to control the formation and fragmentation of vortices in single- and multi-pair production.
It may offer a route to probing topological changes in vacuum decay, although the required intensities remain far above those available at current laser facilities.

\begin{acknowledgments}
The work is supported by the National Natural Science Foundation of China (NSFC) under Grant Nos. 11974419 and 11705278, and by the Fundamental Research Funds for the Central Universities (No. 2023ZKPYL02, 2025 Basic Sciences Initiative in Mathematics and Physics).
\end{acknowledgments}

\end{document}